\input{aipcheck.tex}

\def\selectedoptions{}
\ifx\empty\selectedoptions
  \def\selectedoptions{final}
\fi

\documentclass[
   \selectedoptions
  ]
  {aipproc}

\def\selectedlayoutstyle{6x9} 

\def\kpnnp{$K^+ \to \pi^+ \nu\bar\nu$}
\def\bkpnnp{$B(K^+ \to \pi^+ \nu\bar\nu)$}
\def\kpnn0{$K_L\to\pi^0\nu\bar\nu$}
\def\bkpnn0{$B(K_L\to\pi^0\nu\bar\nu)$}
\def\kmm{$K_L\to\mu^+\mu^-$}
\def\bkmm{$B(K_L\to\mu^+\mu^-)$}
\def\bkmmr{B(K_L\to\mu^+\mu^-)}
\def\kp0{$K_L \to \pi^0 \pi^0$}

\def\bkppr{B(K_L \to \pi^+ \pi^-)}
\def\kpll{$K_L \to \pi^0 \ell^+ \ell^-$}
\def\kpmm{$K_L \to \pi^0 \mu^+ \mu^-$}
\def\kpgg{$K_L \to \pi^0 \gamma\gamma$}

\def\be{\begin{equation}}
\def\ee{\end{equation}}
\def\bea{\begin{eqnarray}}
\def\eea{\end{eqnarray}}

\layoutstyle\selectedlayoutstyle

\SetInternalRegister\hbadness{8000} 

\newcommand\doingARLO[2][]{%
  \ifx\mmref\undefined #1\else #2\fi
}

\begin{document}

\title 
      []
      {Rare K Decays: Results and Prospects}


\author{Laurence Littenberg}{
  address={Brookhaven National Laboratory, Upton, NY 11973},
  email={littenbe@bnl.gov}
}

\copyrightyear  {2001}

\begin{abstract}
Recent results on rare kaon decays are reviewed and prospects for future
experiments are discussed.
\end{abstract}

\date{\today}

\maketitle

\section{Introduction}

In recent years the study of the rare decays of kaons has had three primary
motivations. The first is the search for physics beyond the Standard
Model (BSM).  Virtually all attempts to redress the theoretical
shortcomings of the Standard Model (SM) predict some degree of lepton flavor
violation (LFV).  Decays such as $K_L \to \mu^{\pm} e^{\mp}$ have very
good experimental signatures and can consequently be pursued to remarkable
sensitivities.  These sensitivities correspond to extremely high energy scales
in models where the only suppression is that of the mass of the
exchanged field.  There are also theories that predict new particles
created in kaon decay or the violation of symmetries other than lepton
flavor.

        The second is the potential of decays that are allowed 
but that are extremely suppressed in the SM.  In several of 
these, the leading component
is a G.I.M.-suppressed\cite{Reference:GIM} one-loop process that 
is quite sensitive to fundamental SM parameters such as $V_{td}$.
These decays are also potentially very sensitive to BSM physics.

           Finally there are a number of long-distance-dominated
decays which can test theoretical techniques such
as chiral Lagrangians that purport to explain the low-energy behavior of
QCD.  Knowledge of some of these decays is also needed to extract
more fundamental information from certain of the one-loop processes.

        This field is quite active as indicated by Table \ref{decays:a} 
that lists the
decays for which results have been forthcoming in the last couple of years
as well as  those that are under analysis.  Thus in a short review
such as this,  one must be quite selective.

\begin{table}[h]
\begin{tabular}{llll} 
\hline
$K^+ \to \pi^+ \nu\bar\nu$ & $K_L \to \pi^0 \nu\bar\nu$ &
$K_L \to \pi^0 \mu^+\mu^-$ & $K_L \to \pi^0 e^+e^-$ \\
$K^+ \to \pi^+ \mu^+\mu^-$ & $K^+ \to \pi^+ e^+e^-$ &
$K_L \to  \mu^+\mu^-$ & $K_L \to  e^+e^-$ \\
$K^+ \to \pi^+ \pi^0 \nu\bar\nu$ & $K^+ \to \pi^+ e^+ e^- \gamma$ & 
$K^+ \to \pi^+ \gamma \gamma$ & $K_L \to \pi^0 \gamma \gamma$ \\
$K_L \to \pi^0 e^+ e^- \gamma$ & $K^+ \to \pi^+ \pi^0 \gamma$ & 
$K_L \to \pi^+ \pi^- \gamma$ & $K^+ \to \pi^+ \pi^0 e^+ e^-$ \\
$K_L \to \pi^+ \pi^- e^+ e^-$ & $K^+ \to \mu^+ \nu \gamma$ & 
$K^+ \to \mu^+ \nu e^+ e^-$ & $K^+ \to e^+ \nu e^+ e^-$ \\ 
$K^+ \to e^+ \nu \mu^+\mu^-$ & $K_L \to e^+ e^- \gamma$ & 
$K_L \to \mu^+ \mu^- \gamma$ & $K_L \to e^+ e^- e^+ e^-$ \\
$K_L \to e^{\pm} e^{\mp} \mu^{\pm} \mu^{\mp}$ & $K_L \to e^+ e^- \gamma\gamma$ & 
$K_L \to \mu^+ \mu^- \gamma\gamma$  & $K^+ \to \pi^0 \mu^+ \nu \gamma$ \\
$K^+ \to \pi^+ \mu^+e^-$ & $K_L \to \pi^0 \mu^{\pm} e^{\mp}$ &
$K_L \to \mu^{\pm} e^{\mp}$ & $K^+ \to \pi^- \mu^+ e^+$ \\
$K^+ \to \pi^- e^+ e^+$ & $K^+ \to \pi^- \mu^+ \mu^+$ &
$K^+ \to \pi^+ X^0$ & $K_L \to e^{\pm} e^{\pm} \mu^{\mp} \mu^{\mp}$ \\
$K^+ \to \pi^+ \gamma$ &&& \\
\hline
\end{tabular}
\caption{\it Rare $K$ decay modes under recent or on-going study.}
\label{decays:a}
\end{table}

\section{Beyond the Standard Model}

        There were several $K$ decay experiments dedicated to 
lepton flavor violation at the Brookhaven AGS during the
1990's.  These advanced the sensitivity to such processes
by many orders of magnitude.  In addition, several
``by-product'' results on LFV and other BSM topics 
have emerged from the 
other kaon decay experiments of this period.
Table~\ref{BSM} summarizes the status of work on BSM probes
in kaon decay.

\begin{table}[h]
\begin{tabular}{|c|c|c|c|c|} \hline
Process & Violates & Limit & Experiment & Reference \\
\hline
\hline
$K_L \to \mu e $ & LF& $4.7 \times 10^{-12}$ & AGS-871 & \cite{Ambrose:1998us} \\
$K^+ \to \pi^+ \mu^+ e^-$ & LF &  $2.8 \times 10^{-11}$ & AGS-865 &\cite{Appel:2000wg}  \\
$K^+ \to \pi^+ \mu^- e^+$ & LF, G &$5.2 \times 10^{-10}$ & AGS-865 & \cite{Appel:2000tc} \\
$K_L \to \pi^0 \mu e$ & LF & $4.4 \times 10^{-10}$ & KTeV & \cite{Ledovskoy}\\
$K^+ \to \pi^- e^+ e^+$ & LN, G &$6.4 \times 10^{-10}$ & AGS-865 & \cite{Appel:2000tc} \\
$K^+ \to \pi^- \mu^+ \mu^+$ & LN, G &$3.0 \times 10^{-9}$ & AGS-865 & \cite{Appel:2000tc} \\
$K^+ \to \pi^- \mu^+ e^+$ & LF, LN, G &$5.0 \times 10^{-10}$ & AGS-865 & \cite{Appel:2000tc} \\
$K_L \to \mu^{\pm} \mu^{\pm} e^{\mp} e^{\mp} $ & LF, LN, G & $1.36 \times 10^{-10}$ & KTeV & \cite{Alavi-Harati:2001tk}\\
$K^+ \to \pi^+ f^0$ & N & $5.9 \times 10^{-11}$& AGS-787 & \cite{Adler:2001xv}\\
$K^+ \to \pi^+ \gamma$ & H & $3.6 \times 10^{-7}$& AGS-787 & \cite{Adler:2001dt}\\
\hline
\end{tabular}
\caption{\it Current 90\% CL limits on  $K$ decay modes violating the SM.  The violation
codes are ``LF'' for lepton flavor, ``LN'' for lepton number,
``G'' for generation number, \cite{Cahn:1980kv}, ``H'' for helicity, ``N'' requires new particle}
\label{BSM}
\end{table}

It is clear from this table that any deviation from the SM must be
highly suppressed.  The LFV probes in particular have become the
victims of their own success.  The specific theories they were
designed to test have been killed or at least forced to retreat to the
point where meaningful tests in the kaon system would be very
difficult.  Both kaon flux and rejection of background are becoming
problematical.  Analysis of data already collected is continuing but no
new kaon experiments focussed on LFV are being planned.  Interest in probing
LFV has migrated to the muon sector.

\section{One loop decays}

\begin{figure}[t]
 \includegraphics[angle=90, height=.3\textheight]{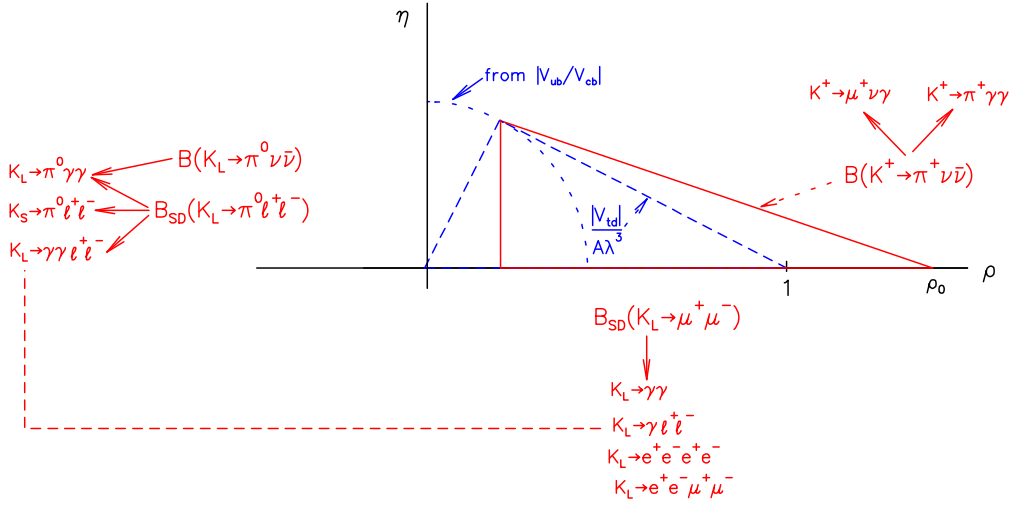}
  \caption{
$K$ decays and the unitarity plane.  The usual unitarity triangle
is dashed.  The triangle that can be constructed from rare $K$ decays
is solid.  See text for further details.
    \label{fig:triangle} }
\end{figure}

In the kaon sector experimental effort has shifted from LFV to
``one-loop'' decays.  These are GIM-suppressed decays in which loops
containing weak bosons and heavy quarks dominate or at least
contribute measurably to the rate.  These processes include \kpnn0,
\kpnnp, \kmm, $K_L \to \pi^0 e^+ e^-$ and $K_L \to \pi^0\mu^+\mu^-$.
In some cases the one-loop contributions violate CP. In one, \kpnn0,
this contribution completely dominates the decay\cite{Littenberg:1989ix}.  
Since the GIM-mechanism tends to enhance the contribution of heavy quarks 
in the loops, in the SM these decays are sensitive to the product of couplings
$V_{ts}^* V_{td}$, often abbreviated as $\lambda_t$. Although one can 
readily analyze these decays in terms of the real and imaginary parts 
of $\lambda_t$, for comparison with results in the $B$ system,
it is conventional to parameterize them in terms of the Wolfenstein variables,
$A$, $\rho$, and $\eta$.  Fig.~\ref{fig:triangle} shows the relationship of
rare kaon decays to the unitarity triangle construction.  The dashed triangle 
is the usual one derived from $V^*_{ub} V_{ud} + V^*_{cb} V_{cd} +
V^*_{tb} V_{td}  = 0$, whereas the solid triangle illustrates the information
available from rare kaon decays.   Note that the ``unitarity point'' at the 
apex, $(\rho, \eta)$, can be determined from either triangle, and
disagreement between the $K$ and $B$ determinations implies physics beyond 
the SM. In Fig.~\ref{fig:triangle} 
the branching ratio closest to each side of the triangle
determines the length of that side.  The arrows leading outward from those
branching ratios point to processes that need to be studied either because
they potentially constitute backgrounds, or because knowledge of them is
required to relate the innermost branching ratios to fundamental parameters.
$K_L \to \mu^+ \mu^-$, which can determine the bottom of the triangle ($\rho$),
is the process for which the experimental data is the best, but for which
the theory is most problematical.  \kpnn0, which determines
the height of the triangle is theoretically the cleanest, but experiment is
many orders of magnitude short of the SM-predicted level.  \kpnnp,
which determines the hypotenuse, is nearly as clean as \kpnn0~
and has been observed. 
Prospects for \kpnnp~ are probably the best of the three since it
is already clear it can be exploited.

\subsection{$K_L \to \mu^+ \mu^-$}

\begin{figure}[t]
 \includegraphics[angle=0, height=.25\textheight]{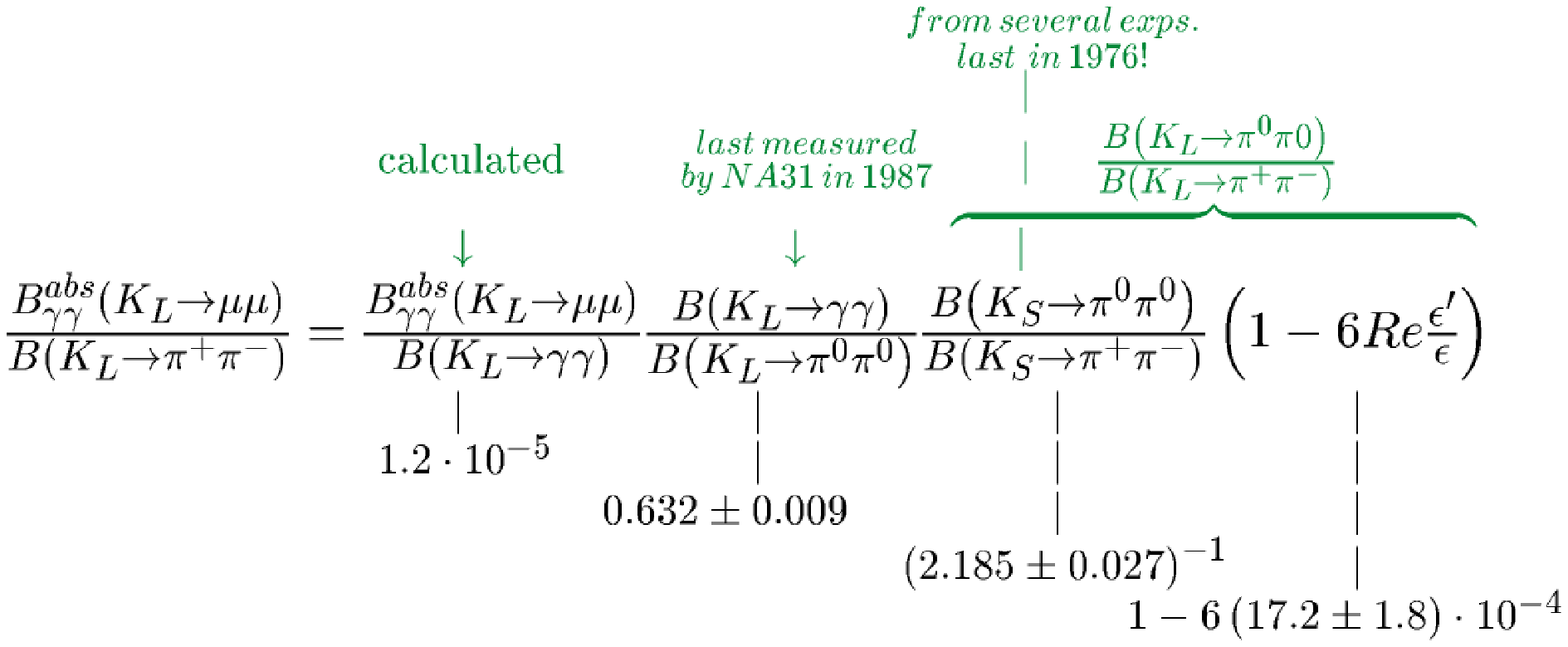}
  \caption{Components of the calculation of 
$\frac{B^{abs}_{\gamma\gamma}(K_L \to \mu^+\mu^-)}{\bkppr}$.
    \label{fig:mu2} }
\end{figure}

The short distance component of this decay can be  quite reliably
calculated in the SM\cite{Buchalla:1994wq}.  The most recent
measurement of its branching ratio\cite{Ambrose:2000gj} based on some 6200
events gave $B(K_L \to \mu^+ \mu^-) = (7.18 \pm 0.17)\times 10^{-9}$.
However \kmm~ is dominated by long distance effects, the
largest of which, the absorptive contribution mediated by $K_L \to
\gamma\gamma$, accounts for $(7.07 \pm 0.18)\times 10^{-9}$.
Subtracting the two, yields a 90\% CL upper limit on the total
dispersive part of \bkmm~ of $0.37 \times 10^{-9}$.  One can do a little 
better than this in the following way.  The actual quantity measured in 
Ref~\cite{Ambrose:2000gj} was $\frac{\bkmmr}{\bkppr} = (3.48 \pm 0.05)\times 
10^{-6}$  One wants to subtract from this measured quantity the ratio 
$\frac{B^{abs}_{\gamma\gamma}(K_L \to \mu^+\mu^-)} {\bkppr}$.  
Fig~\ref{fig:mu2} shows the components of this latter ratio,
obtained from Ref.~\cite{Groom:2000in}, whose
product is $(3.435 \pm 0.065) \times 10^{-6}$.  

The subtraction yields $\frac{B^{disp}(K_L \to \mu^+ \mu^-)}{\bkppr} = 
(0.045 \pm 0.082) \times 10^{-6}$ (where $B^{disp}$ refers to the 
dispersive part of \bkmm).  $\frac{B^{disp}(K_L \to \mu^+ \mu^-)}{\bkppr}$
can then be
multiplied by $\bkppr = (2.056 \pm 0.033) \times 10^{-3}$ to obtain
$B^{disp}(K_L \to \mu^+\mu^-) = (0.093\pm 0.169) \times 10^{-9}$, or
$B^{disp}(K_L \to \mu^+\mu^-) < 0.31 \times 10^{-9}$ at 90\% CL.  Note
that some of the components represent quite old measurements.  Since
\bkmm~and $B^{abs}_{\gamma\gamma}(K_L \to \mu^+ \mu^-)$ are so close,
small shifts in the component values could have relatively large consequences
for  $B^{disp}(K_L \to \mu^+\mu^-)$.  Several of the components could be
remeasured by experiments presently in progress\footnote{There is a 
new preliminary result from the KLOE experiment of 
$B(K_S \to \pi^+ \pi^-)/B(K_S \to \pi^0 \pi^0) = 2.192 \pm 0.003_{stat} 
\pm 0.016_{syst}$\cite{DiDomenico:2001}. Fortuitously, inserting this result in 
place of the PDG value makes no difference to the final limit.}.  
Now if one inserts
the result of even very conservative recent CKM fits into the formula for 
the short distance part of \bkmm, one gets rather poor agreement with
the limit of $B^{disp}(K_L \to \mu^+\mu^-)$ derived above.  For example the
95\% CL fit of Hocker et al.\cite{Hocker:2001xe}\cite{Hocker:2001jb}, $\bar\rho = 0.07 - 0.37$,
gives $B^{SD}(K_L \to \mu^+\mu^-) = (0.4-1.3)\times 10^{-9}$.  So why 
haven't we been hearing about this apparent violation of the SM?

\begin{figure}[t]
 \includegraphics[angle=0, height=.25\textheight]{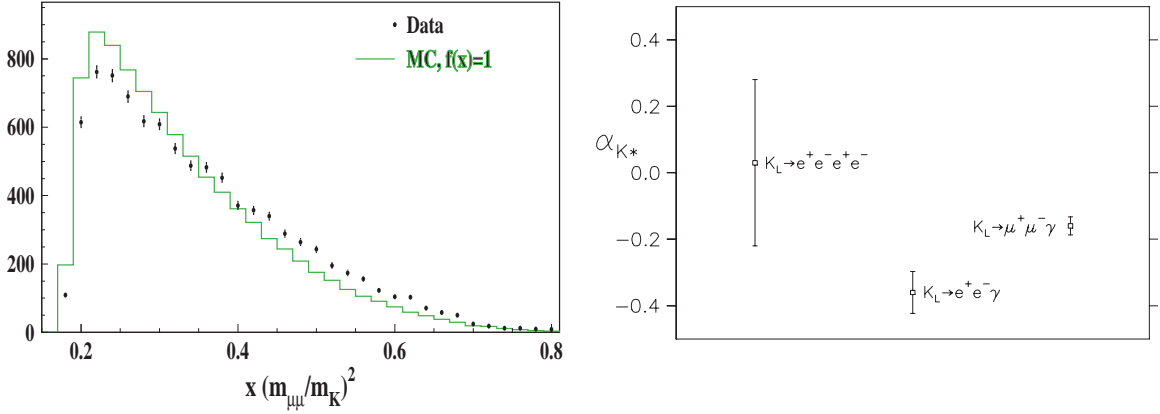}
  \caption{Left: Spectrum of $x = (m_{\mu\mu}/m_K)^2$ in 
$K_L \to \mu^+ \mu^- \gamma$ from Ref.\cite{Alavi-Harati:2001wd}.  
Right: Determinations of the BMS parameter
$\alpha_{K^*}$ from three $K_L$ decays involving virtual photons.
    \label{fig:mmg} }
\end{figure}

The answer is that unfortunately $K_L \to \gamma^* \gamma^* $ also
gives rise to a dispersive contribution, which is much less tractable
than the absorptive part, and which can interfere with the
short-distance weak contribution that one is trying to extract.  The
problem in calculating this contribution is the necessity of including
intermediate states with virtual photons of all effective masses.
Thus such calculations can only partially be validated by studies of
processes containing virtual photons in the final state.  Recently
there have been publications on $K_L \to \gamma \mu^+
\mu^-$\cite{Alavi-Harati:2001wd} (9327 events), $K_L \to e^+ e^- e^+
e^-$\cite{Alavi-Harati:2001ab} (441 events), and $K_L \to \mu^+ \mu^-
e^+ e^-$\cite{Alavi-Harati:2001tk} (38 events) and there exist
slightly older high statistics data on $K_L \to \gamma e^+
e^-$\cite{Fanti:1999rz} (6864 events).  Figure~\ref{fig:mmg}-left
shows the spectrum of $x = (m_{\mu\mu}/m_K)^2$ from
Ref.\cite{Alavi-Harati:2001wd}.  The disagreement between the data
(filled circles with error bars) and the prediction of pointlike
behavior (histogram) clearly indicates the presence of a form factor.  A
long-standing candidate for this is provided by the BMS
model\cite{Bergstrom:1983rj} which depends on a single parameter,
$\alpha_{K^*}$.

Fig.~\ref{fig:mmg}-right shows three determinations of this parameter.
The level of agreement of these results leaves something to be
desired.  Fitting to a more recent parameterization of these
decays\cite{D'Ambrosio:1998jp} also results in quite marginal
agreement.  This may improve when radiative corrections are
properly taken into account.  Thus additional effort, both
experimental and theoretical, is required before the quite precise data on
\bkmm~ can be fully exploited.

\subsection{\kpnnp}

\begin{figure}[t]
 \includegraphics[angle=0, height=.25\textheight]{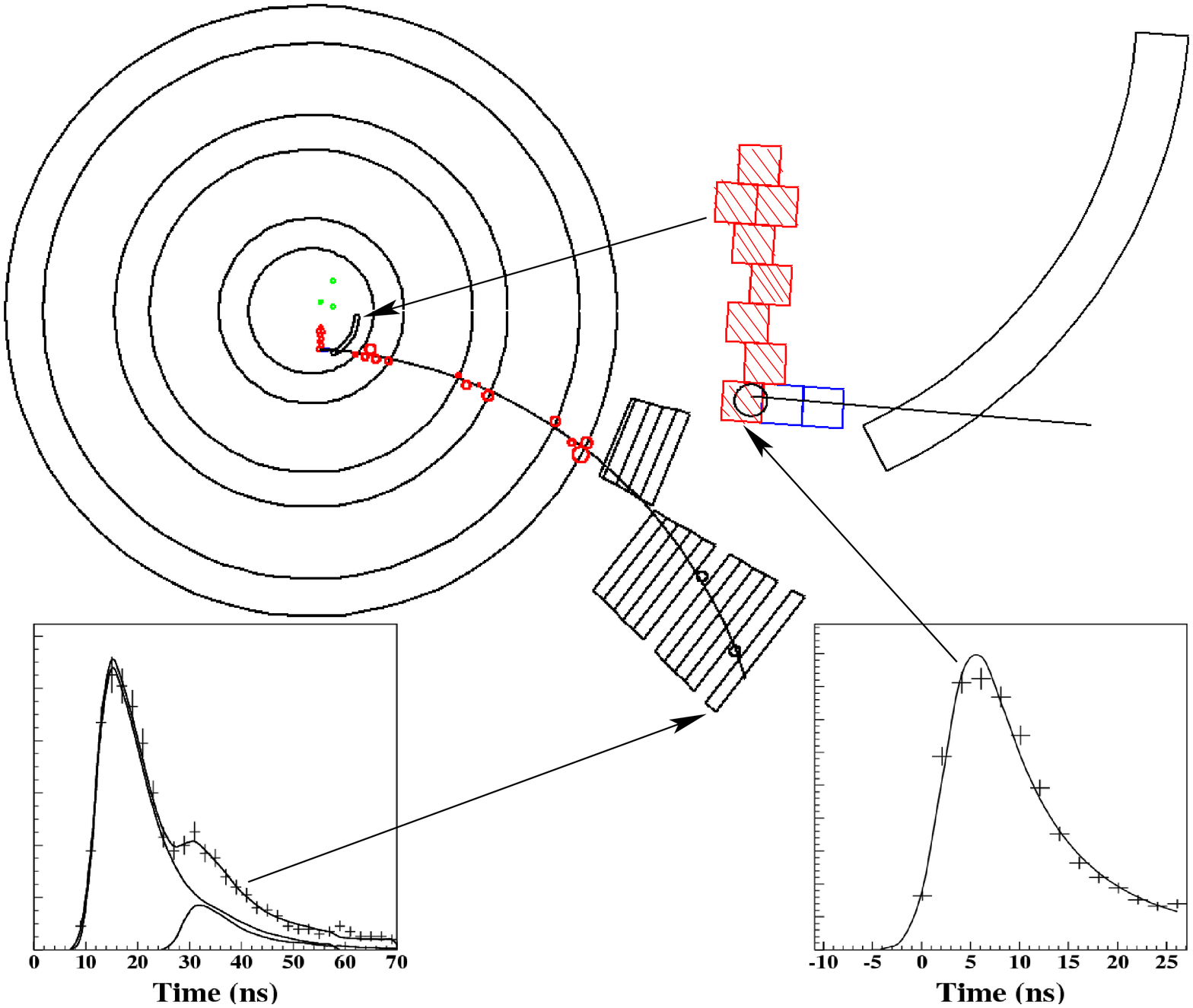}
\includegraphics[angle=0, height=.25\textheight]{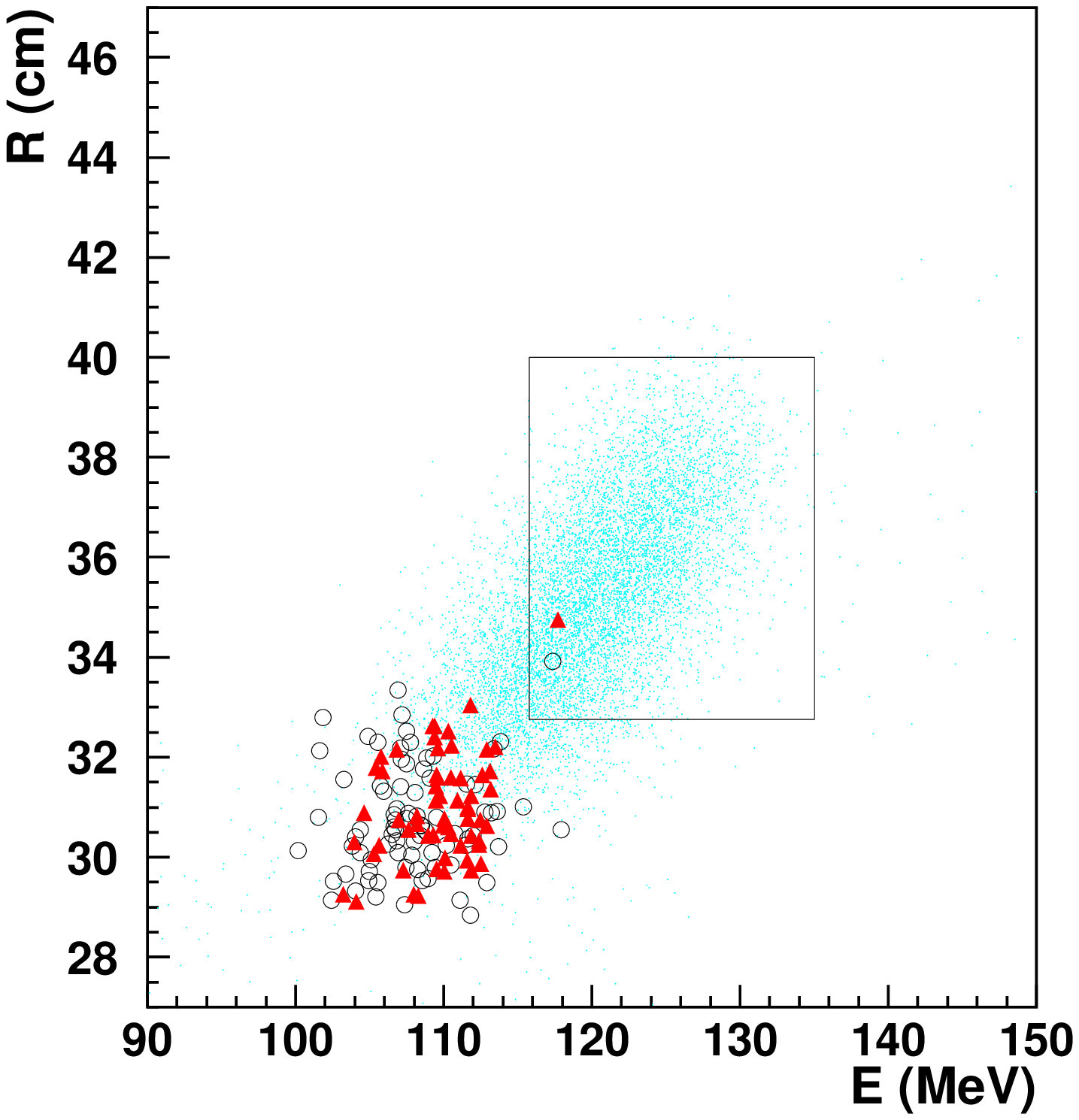}
  \caption{
Left: new \kpnnp~event. Right: Range vs energy of $\pi^+$ in the final
sample.  The circles are 1998 data and the triangles 1995-7 data.  The events 
around $E=108$ MeV are $K^+ \to \pi^+ \pi^0$ background.  The simulated
distribution of expected signal events is indicated by dots.
    \label{fig:e787} }
\end{figure}

Theoretically \kpnnp~ is remarkably clean, suffering from none of the 
long distance complications of $K_L \to \mu^+ \mu^-$.  
The hadronic matrix element, so often a problem in other processes,
can be calculated to a $\sim 2\%$ via an isospin
transformation from that of $K_{e3}$\cite{Marciano:1996wy}.
Interest in \kpnnp~ is driven in large part by its
sensitivity to $V_{td}$ (it is actually directly sensitive to the quantity
$|V_{ts}^* V_{td}|$).  Its amplitude is proportional to
the dark slanted line at the right in Fig.~\ref{fig:triangle}.  This is
equal to the vector sum of the line proportional to $|V_{td}|/A \lambda^3$
(where $\lambda \equiv sin \theta_{Cabibbo}$)
and that from $(1,0)$ to the point marked $\rho_0$. The length $\rho_0 -1$ 
along the real
axis is proportional to the amplitude for the charm contribution to
\kpnnp.  The QCD corrections to this amplitude, which
are responsible for the largest uncertainty in 
\bkpnnp, have been calculated to NLLA\cite{Buchalla:1994wq}. The residual
uncertainty in the charm amplitude is estimated to be $\sim 15\%$ which
leads to only  a $\sim 6\%$ uncertainty\cite{Buchalla:1998ba} in
extracting $|V_{td}|$ from \bkpnnp.  Recently AGS E787 has seen evidence for
a second event of \kpnnp~\cite{Adler:2001xv} (see Fig.~\ref{fig:e787})
which, combined with previous data \cite{Adler:2000by},
yields a branching ratio \bkpnnp $= (1.57 {+1.75 \atop -0.82}) \times 
10^{-10}$. By comparison, a fit to the CKM phenomenology yields the expectation
$(0.72 \pm 0.21) \times 10^{-10}$\cite{D'Ambrosio:2001zh}.
It is notable that E787 has established methods to reduce the
residual background to $\sim 10\%$ of the the signal branching ratio
predicted by the SM.

A new experiment, AGS E949\cite{e949}, based on an upgrade of the
E787 detector, is about to begin its first physics run.
Using the entire flux of the AGS for 6000 hours, it is designed to
reach a sensitivity of $\sim 10^{-11}$/event.  In June 2001, Fermilab 
gave Stage 1 approval to an
experiment (CKM\cite{Cooper:2001ba}) to extend the study of \kpnnp~
by yet an another order of magnitude in sensitivity.  This experiment,
unlike all previous ones on this process, uses an in-flight rather
than a stopping $K^+$ technique.  This experiment is expected to
start collecting data in 2007 or 2008.

\begin{figure}[h]
 \includegraphics[angle=0, height=.4\textheight]{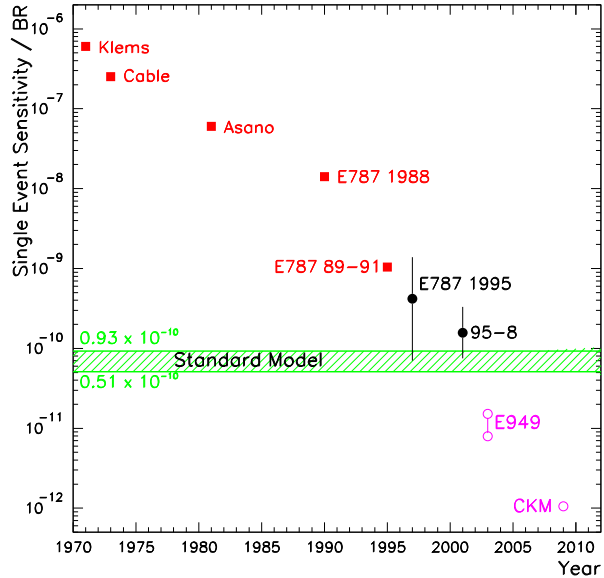}
  \caption{History and prospects for the study of \kpnnp.  Points without
error bars are single event sensitivities, those with error bars are 
measured branching ratio.
    \label{fig:prog} }
\end{figure}

Fig.~\ref{fig:prog} shows the history and expectations of progress in
studying \kpnnp.

\subsection{\kpnn0}
\kpnn0 is the most attractive target in the kaon system, since it is
direct CP-violating to a very good
approximation\cite{Littenberg:1989ix,Buchalla:1998ux} (\bkpnn0 $\propto
\eta^2$).  Like \kpnnp\ it has a
hadronic matrix element that can be obtained from $K_{e3}$, but,
it has no significant contribution from charm.  As a result, the
intrinsic theoretical uncertainty connecting  \bkpnn0\ to
the fundamental SM parameters is only about 2\%.  
Note also that \bkpnn0~
is directly proportional to the square of $Im \lambda_t$ and
that $Im \lambda_t = - $$\cal{J}$$/[\lambda (1-\frac{\lambda^2}{2})]$
where $\cal{J}$ is the Jarlskog invariant\cite{Jarlskog:1985ht}.
Thus a measurement of \bkpnn0~determines the area of the unitarity
triangles with a precision twice as good as that on \bkpnn0~itself.

        \bkpnn0\ can be bounded indirectly by measurements of \bkpnnp\
through a nearly model-independent relationship pointed out by
Grossman and Nir\cite{Grossman:1997sk}.  The application of this to
the new E787 result yields \bkpnn0$<1.7 \times 10^{-9}$ at 90\% CL.
This is far tighter than the current direct experimental limit, $5.9
\times 10^{-7}$, obtained by KTeV\cite{Alavi-Harati:1999hd}.  To
actually measure \bkpnn0 at the SM level ($\sim 3 \times 10^{-11}$),
one will need to improve on this by some five orders of magnitude.
The KEK E391a experiment\cite{Inagaki:1997gc} proposes to achieve a 
sensitivity of $\sim 3 \times 10^{-10}$/event which would better 
the indirect limit by a factor five, but would not quite bridge this gap.
It will serve as a test for a future much more sensitive
experiment to be performed at the Japanese Hadron Facility.  E391a
features a carefully designed ``pencil'' beam, and a very high
performance photon veto.  The active photon detector is a CsI-pure
crystal calorimeter.  The entire rather compact apparatus will operate
in vacuum.  Beamline construction and tuning started in March 2000 and
physics running is expected to begin in Fall, 2003.

\begin{figure}[h]
 \includegraphics[angle=0, height=.52\textheight]{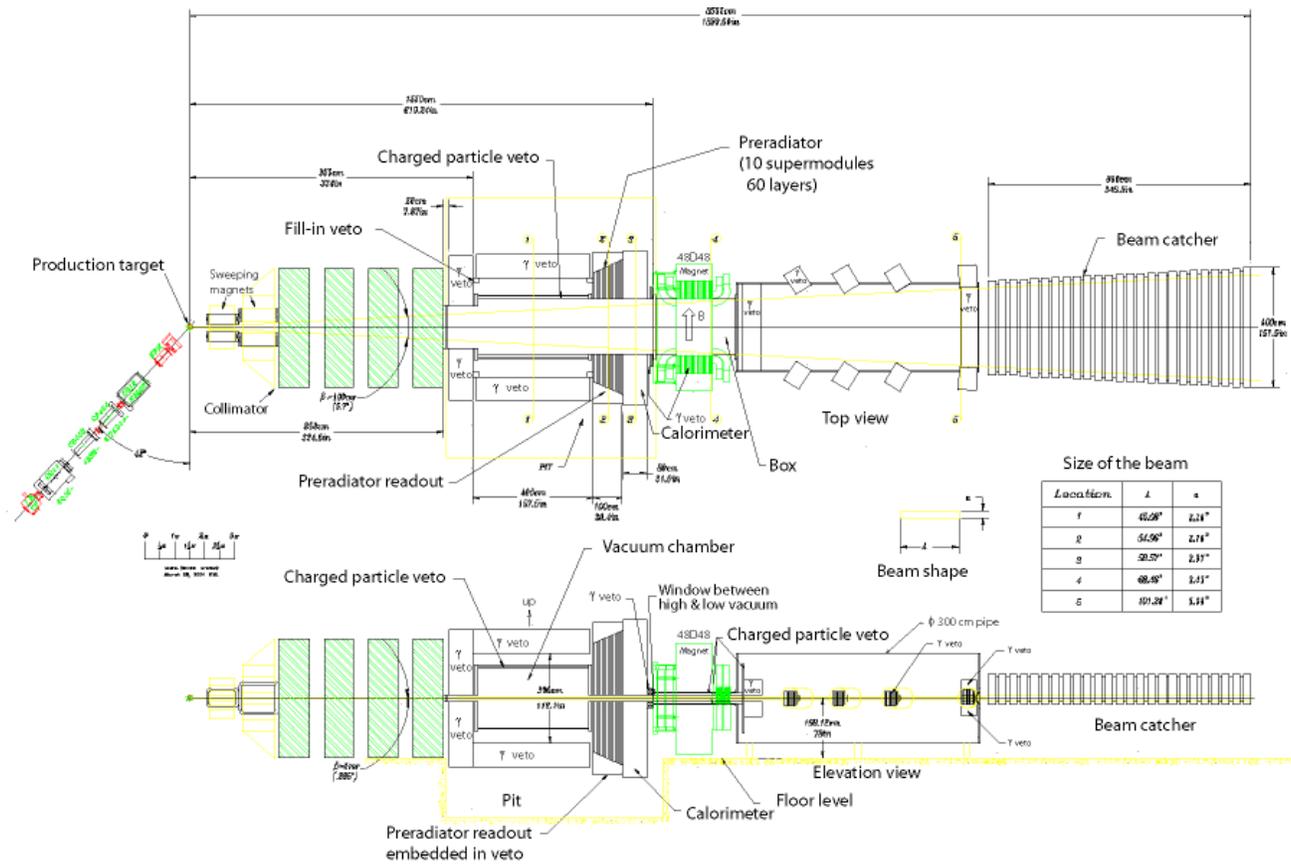}
  \caption{Layout of the KOPIO detector.
    \label{fig:kopio} }
\end{figure}

        The KOPIO experiment\cite{Konaka:1998hk} at BNL (E926) takes
a completely different approach, exploiting the intensity and flexibility of
the AGS to make a high-flux, low-energy, microbunched $K_L$ beam.
The proposed experiment is shown in Fig.~\ref{fig:kopio}.  The neutral
beam will be extracted at $\sim 45^{\rm o}$ to soften the $K_L$ spectrum
sufficiently to permit time-of-flight determination of the $K_L$ velocity.
The large production angle also softens the neutron spectrum 
so that they (and the $K_L$) are by and large below threshold for the
hadro-production of $\pi^0$'s.  The beam region will be evacuated to
$10^{-7}$ Torr to further minimize such production.  With a 10m beam channel 
and this low energy beam, the contribution of hyperons to the background will
be negligible.  $K_L$ decays from a $\sim 3$m fiducial region will be 
accepted.  Signal photons impinge on
a 2 $X_0$ thick preradiator capable of measuring their direction.  
An alternating drift chamber/scintillator plane structure will allow energy 
measurement as well.  A high-precision shashlyk calorimeter downstream of 
the preradiator will complete the energy measurement.  The photon
directional information will allow the decay vertex position to be determined.
Combined with the target position and time of flight information,
this provides a measurement of the $K_L$ 3-momentum so that
kinematic constraints as well as photon vetoing are available to suppress
backgrounds.  The leading expected background is \kp0, which is 
initially some eight orders of magnitude larger than the predicted signal.  
However since $\pi^0$'s from this background have a unique energy in
the $K_L$ center of mass, a very effective kinematic cut can be
applied.  This reduces the load on the photon veto system
surrounding the decay region to the point where the hermetic 
veto techniques proven in E787 are sufficient.  In fact most of
the techniques necessary for KOPIO have been proven in previous 
experiments or in prototype tests.  KOPIO aims
to collect about $50$ \kpnn0\ events with a signal to background
ratio of 2:1.  This will permit $\eta$ to be determined to $\sim 10\%$,
given expected progress in measuring $m_t$ and $V_{cb}$.  KOPIO will
run during the $\sim$20 hours/day the AGS is not needed for injection 
into RHIC.

\subsection{\kpll}

	These are reactions initially thought experimentally more
tractable than \kpnn0.  Like \kpnn0, in the SM they are sensitive to
$Im \lambda_t$, but in general they have different sensitivity to BSM
effects\cite{Buras:1999da}.  Although their signatures are
intrinsically superior to that of \kpnn0, they are subject to a
serious background that has no analogue in the case of the latter: $K_L \to
\gamma\gamma\ell^+\ell^-$.  This process, a radiative correction to
$K_L \to \gamma\ell^+\ell^-$, occurs roughly $10^5$ times
more frequently than \kpll.  Kinematic cuts are
quite effective, but it is very difficult to improve the signal:background
beyond about $1:5$\cite{Greenlee:1990qy}.  Both varieties of  $K_L \to
\gamma\gamma\ell^+\ell^-$ have been observed, $B(K_L \to \gamma\gamma e^+ 
e^-)_{k_{\gamma}>5 MeV} = (5.84 \pm 0.15 (stat) \pm 0.32 (syst)) \times
10^{-7}$\cite{Alavi-Harati:2000tv} and $B(K_L \to \gamma\gamma \mu^+
\mu^-)_{m_{\gamma\gamma}> 1 MeV/c^2} = (10.4 {+7.5 \atop{-5.9}}(stat) \pm 
0.7 (syst)) \times 10^{-9}$\cite{Alavi-Harati:2000hr}; both agree
with theoretical prediction.   By comparison, in the SM 
$B^{direct}(K_L \to \pi^0 e^+ e^-)$ is predicted to be \cite{Buras:2001pn}
$(4.3 \pm 2.1)\times 10^{-12}$ and  $B^{direct}(K_L \to \pi^0 \mu^+ \mu^-)$
about fives times smaller.

	In addition to this background, there are two other issues
that make the extraction of short-distance information from 
\kpll~ problematical.  First, there is an indirect CP-violating
contribution from the $K_1$ component of $K_L$ given by
$|\epsilon|^2 {\tau_{K_L}\over{\tau_{K_S}}} B(K_S \to \pi^0 e^+ e^-)$
which is of the same order of magnitude as the direct CP-violating 
piece\footnote{There is also an interference term between the indirect 
and direct CP-violating amplitudes.}.  The exact size of this
contribution will be predictable if and when $B(K_S \to \pi^0 e^+ e^-)$
is measured, hopefully by the upcoming NA48/1
experiment\cite{NA48-1}.  Second is yet another contribution of
similar size mediated by $K_L \to \pi^0
\gamma\gamma$ which is {\it CP-conserving}.  To some extent 
this contribution can be predicted from measurements of the branching
ratio and kinematic distributions of $K_L \to \pi^0 \gamma\gamma$, and
thousands of these events have been observed.  However as indicated in
Table~\ref{Kpgg}, a new result from NA48\cite{Zinchenko:2001} disagrees by 
nearly $3 \sigma$ from the previous result from 
KTeV\cite{Alavi-Harati:1999mu}.  The change in the vector meson exchange 
contribution, characterized by the parameter $a_V$, reduces the
predicted size of $B^{CP-cons}(K_L \to \pi^0 e^+ e^-)$
considerably\cite{D'Ambrosio:1997sw} which is good news for
the prospects of measuring $B^{direct}(K_L \to \pi^0 e^+ e^-)$.
However the validity of the current technique for predicting
$B^{CP-cons}(K_L \to \pi^0 e^+ e^-)$ from \kpgg~ has recently been 
reexamined \cite{Gabbiani:2001zn}, and questions raised about
the functions used to fit the spectrum and about the treatment of 
the dispersive contribution.  Thus both the theoretical and experimental
situations are quite unsettled at the moment.  Depending on whose
data and whose theory one uses, values from $0.25 \times 10^{-12} $ 
to $7.3 \times 10^{-12}$ are predicted for $B^{CP-cons}(K_L \to \pi^0 e^+ e^-)$.

\begin{table}[h]
\begin{tabular}{llll} \hline
Exp. & $B(K_L \to \pi^0 \gamma\gamma) \times 10^6$ & $a_V$ & Ref. \\
\hline
KTeV & $1.68 \pm 0.07_{stat} \pm 0.08_{syst} $ &
$-0.72 \pm 0.05 \pm 0.06 $ & \cite{Alavi-Harati:1999mu} \\
NA48 & $1.36 \pm 0.03_{stat} \pm 0.03_{syst} \pm 0.03_{norm}$ &
$-0.46 \pm 0.03 \pm 0.03 \pm 0.02_{theor} $ & \cite{Zinchenko:2001} \\
\hline
\end{tabular}
\caption{\it Results on $K_L \to \pi^0 \gamma\gamma$.}
\label{Kpgg}
\end{table}

	The current experimental status of \kpll~is summarized in Table
~\ref{Kpll}.  A factor $\sim 2.5$ more data is expected from the KTeV
1999 run, but as can be seen from the table, background is already
starting to be observed at a sensitivity roughly 100 times short of
the expected signal level.  

\begin{table}[h]
\begin{tabular}{lllll} \hline
Mode & 90\% CL upper limit & Est. bkgnd.& Obs. evts. &Ref. \\
\hline
$K_L \to \pi^0 e^+ e^-$     & $5.1 \times 10^{-10}$ & $1.06\pm 0.41$ & 2 & \cite{Alavi-Harati:2000sk} \\
$K_L \to \pi^0 \mu^+ \mu^-$ & $3.8 \times 10^{-10}$ & $0.87\pm 0.15$ & 2 & \cite{Alavi-Harati:2000hs} \\
\end{tabular}
\caption{\it Results on \kpll.}
\label{Kpll}
\end{table}

\begin{figure}[t]
 \includegraphics[angle=0, height=.25\textheight]{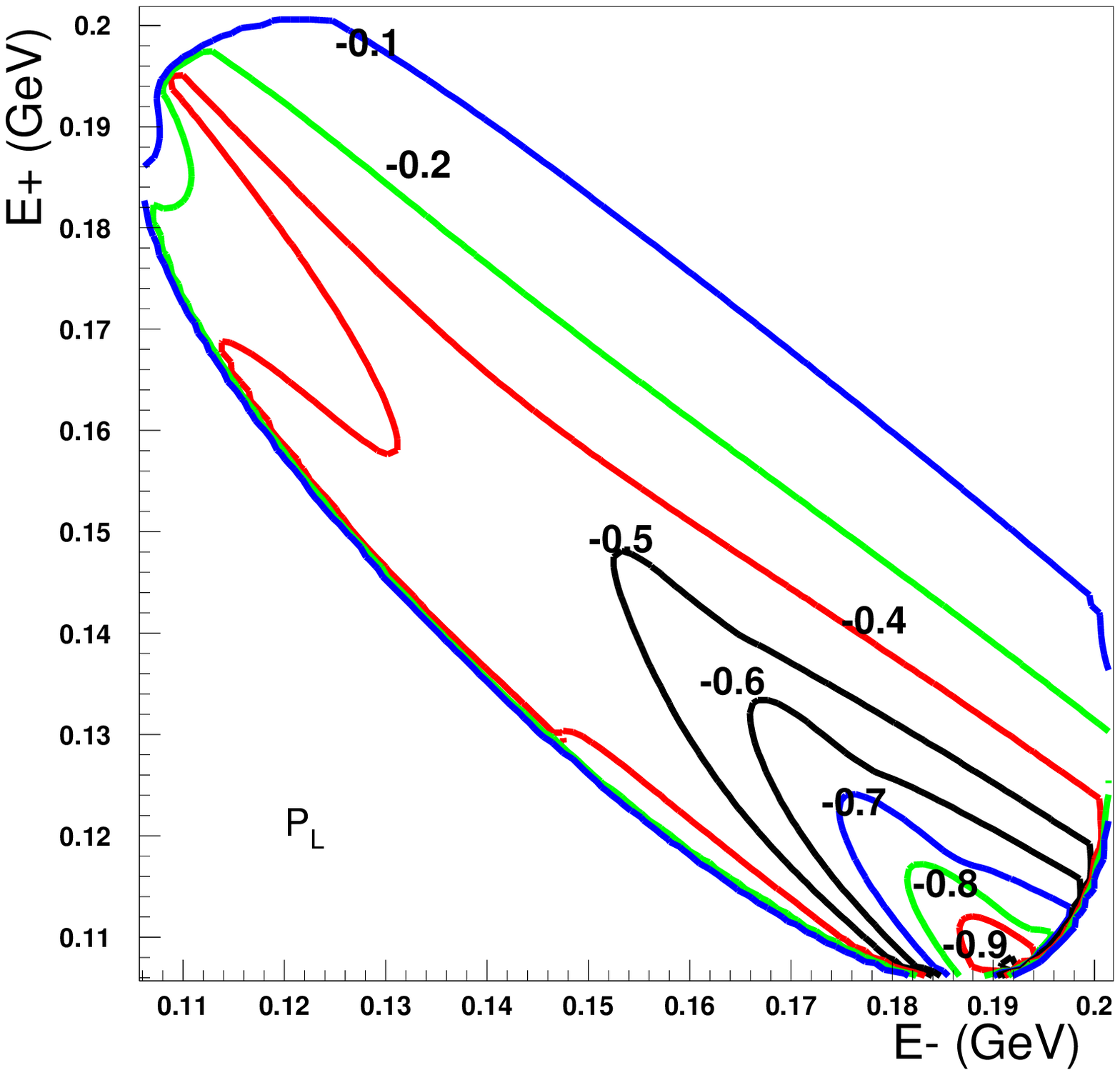}
\includegraphics[angle=0, height=.25\textheight]{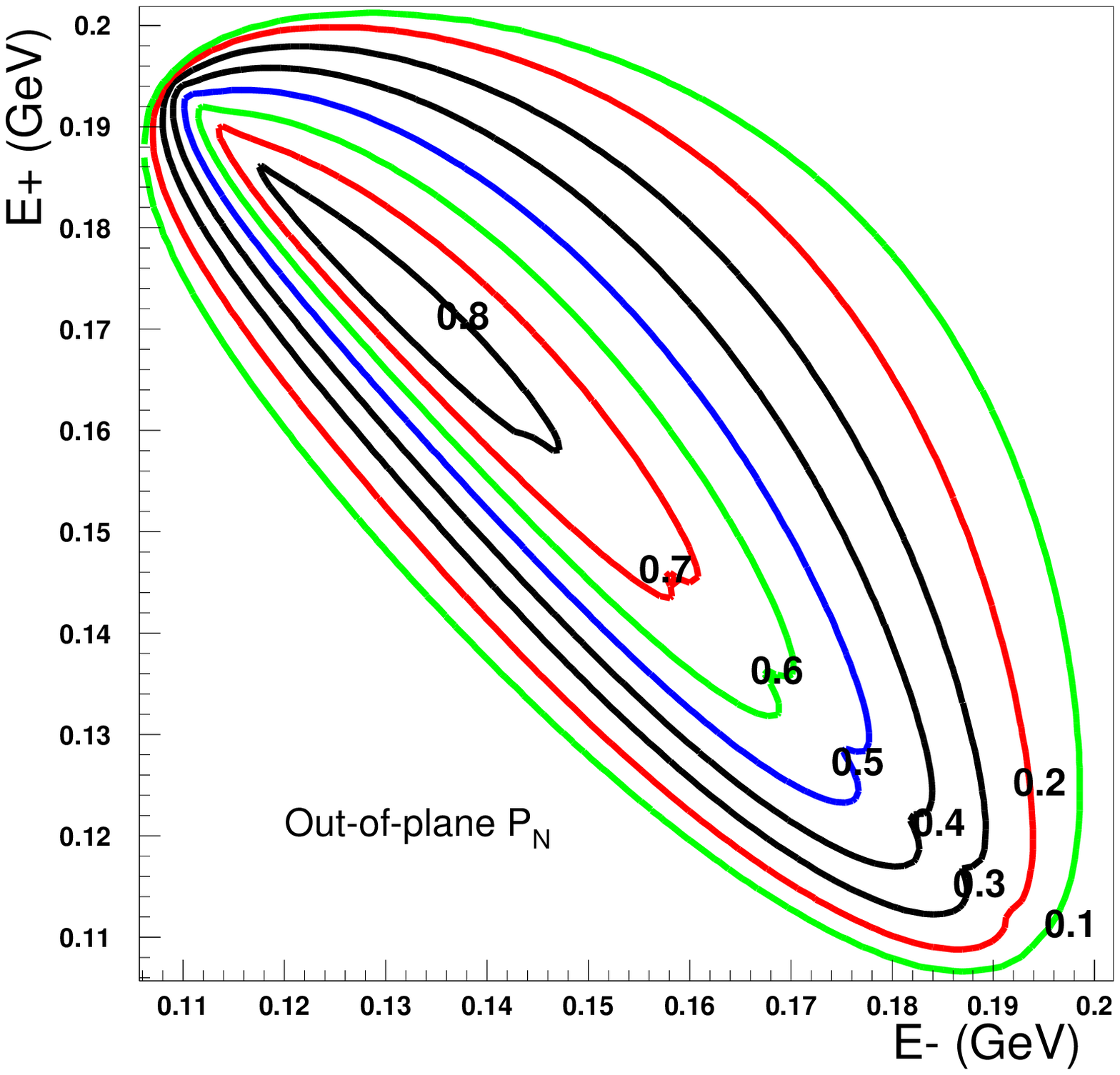}
  \caption{$\mu^+$ polarizations in \kpmm, plotted against the muon cm energies.
Left: longitudinal polarization. Right: out-of-plane polarization.
    \label{fig:md} }
\end{figure}

	To make a useful measurement under these conditions will require
markedly increased statistics on the signal and both theoretical and
experimental advances in the ancillary modes \kpgg~and $K_S \to \pi^0
e^+ e^-$.  Various approaches for mitigating these problems have been
suggested over the years including studies of the Dalitz Plot
\cite{Donoghue:1987aw}, the time development \cite{Kohler:1995rb},
or both \cite{Littenberg:1988cy}.  However an innovative approach has
recently been suggested \cite{Diwan:2001} in which muon polarization
in $K_L \to \pi^0 \mu^+ \mu^-$ as well as kinematic distributions are
exploited.  The $\mu^+$ longitudinal polarization is proportional to the
direct CP-violating amplitude, whereas the energy asymmetry and the
out-of-plane polarization depend on both indirect and direct CP
violating amplitudes.  As shown in Fig.~\ref{fig:md},
the polarizations involved turn out to be
extremely large so that enormous numbers of events may not be
required.

\subsection{An alternative parameterization}

	Although it is customary to write the branching ratios and other
observables of the one-loop processes in terms of the Wolfenstein
parameterization of the CKM matrix, this parameterization is not
really natural to the kaon system, and puts results from this system
at a certain disadvantage in comparisons with those from the $B$ system.
To extract information on $\rho$ and $\eta$, for example, 
it is necessary to divide the physical measurements by
$\lambda^8 A^4$, thereby introducing
``external'' contributions to the  uncertainty  of $8 \sigma_{\lambda}$ and
$4 \sigma_A$.  One can avoid this by resorting 
to expressions for the branching ratios in terms of
the quantity $\lambda_t$.   Since as noted above, the imaginary part of
this quantity determines the area of all unitarity triangles, it is no
less fundamental than $\rho$ and $\eta$.  

	The formulae for the branching ratios of three of the decays
discussed above are:

\begin{eqnarray}\label{brsimp}
B(K^+ \to \pi^+ \nu\bar\nu)& = &\xi [ (\lambda_c \bar X + Re(\lambda_t) X_t)^2  + (Im(\lambda_t) X_t)^2]\\
B^{SD}(K_L \to \mu^+\mu^-)& = &\xi' [Re(\lambda_c) Y_{NL} + Re(\lambda_t) Y(x_t)]^2 \\
B(K_L \to \pi^0 \nu\bar\nu) & = &\xi'' (Im(\lambda_t) X_t)^2
\end{eqnarray}

where 

\begin{eqnarray}\label{xi}
\xi   \equiv  {{3 r_{K^+} \alpha^2 B_{K^+e3}}\over{V_{us}^2 2
\pi^2 sin^4{\theta_W}}} & = & 1.55 \times 10^{-4} \\
\xi'   \equiv  {{ \tau_{K_L} \alpha^2 B_{K^+\mu\nu}}\over{\tau_{K^+} V_{us}^2
\pi^2 sin^4{\theta_W}}} & = & 6.32 \times 10^{-3} \\
\xi''   \equiv  {{3 r_{K_L} \tau_{K_L} \alpha^2 B_{K^+e3}}\over{\tau_{K^+} V_{us}^2 2
\pi^2 sin^4{\theta_W}}} & = & 6.77 \times 10^{-4} 
\end{eqnarray}

\noindent
and to a good approximation the Inami-Lim \cite{Inami:1981fz}
functions $X_t = 1.56 (m_t/170{\rm GeV})^{1.15}$, $Y_t = 1.02
(m_t/170{\rm GeV})^{1.56}$. The quantities $r_{K^+} = 0.901$ and 
$r_{K_L} = 0.944$
are isospin correction factors that relate the hadronic matrix
elements of the $K \to \pi\nu\bar\nu$ processes to that of $K^+ \to
\pi^0 e^+ \nu$ \cite{Marciano:1996wy}.  The terms $\bar X$ and $Y_{NL}$
are the Inami-Lim functions for the charm contributions which, after
correction to NLLA, are known to about 15\%.

\begin{figure}[h]
 \includegraphics[angle=0, height=.25\textheight]{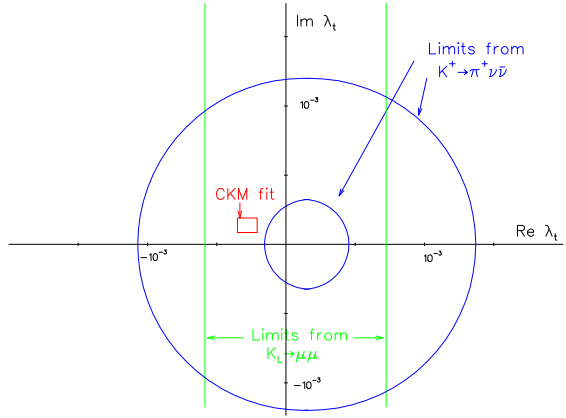}
  \caption{Comparison of 90\% CL constraints from current data on 
rare kaon decays with 95\% CL constraints from a typical unitarity 
fit (based on Ref. \cite{Hocker:2001jb}).  The allowed region 
from kaon decays lies between the two circles and within the outer two 
vertical lines. 
    \label{fig:ut90} }
\end{figure}

\begin{figure}[h]
\includegraphics[angle=0, height=.25\textheight]{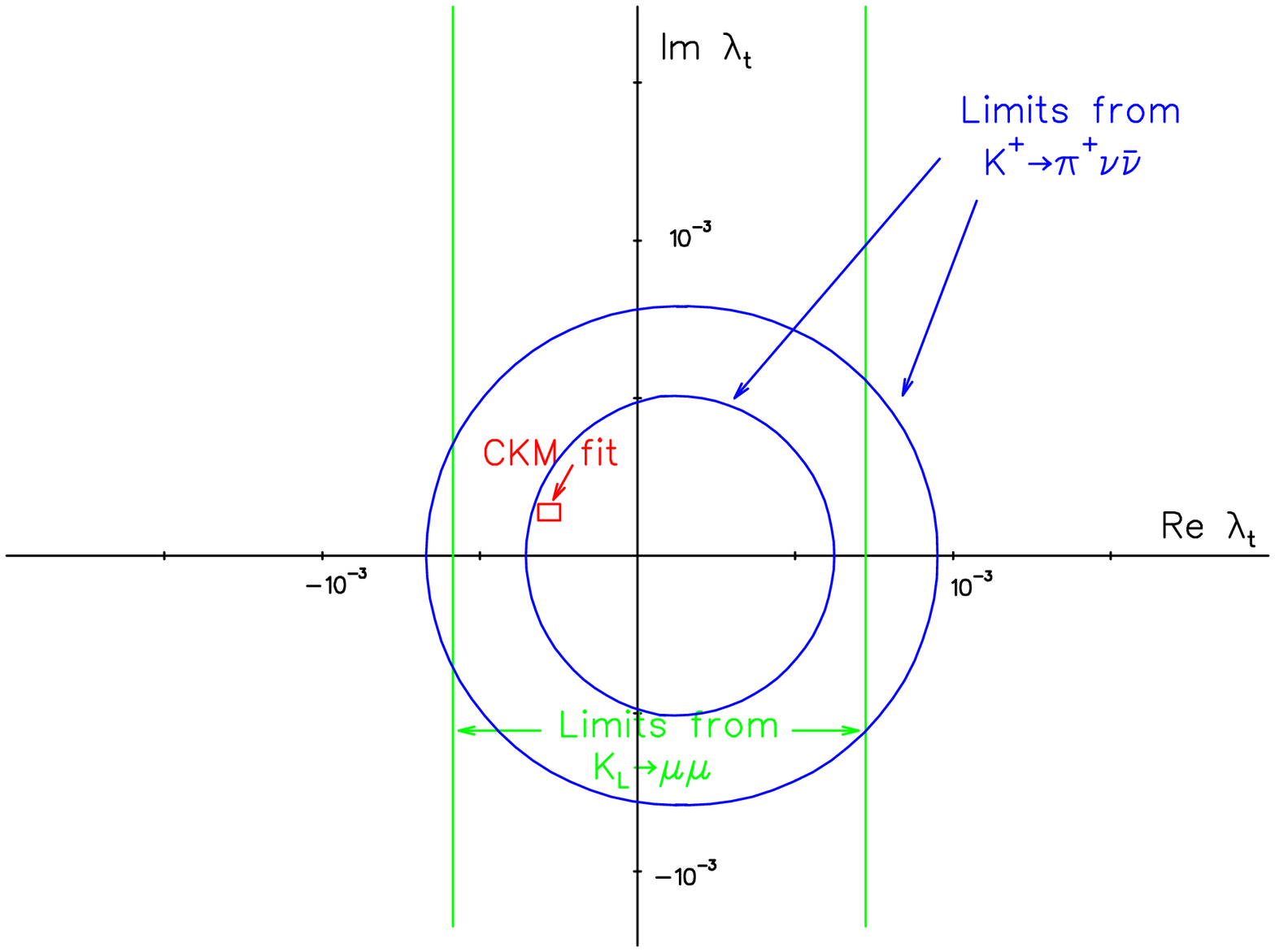}
\includegraphics[angle=0, height=.25\textheight]{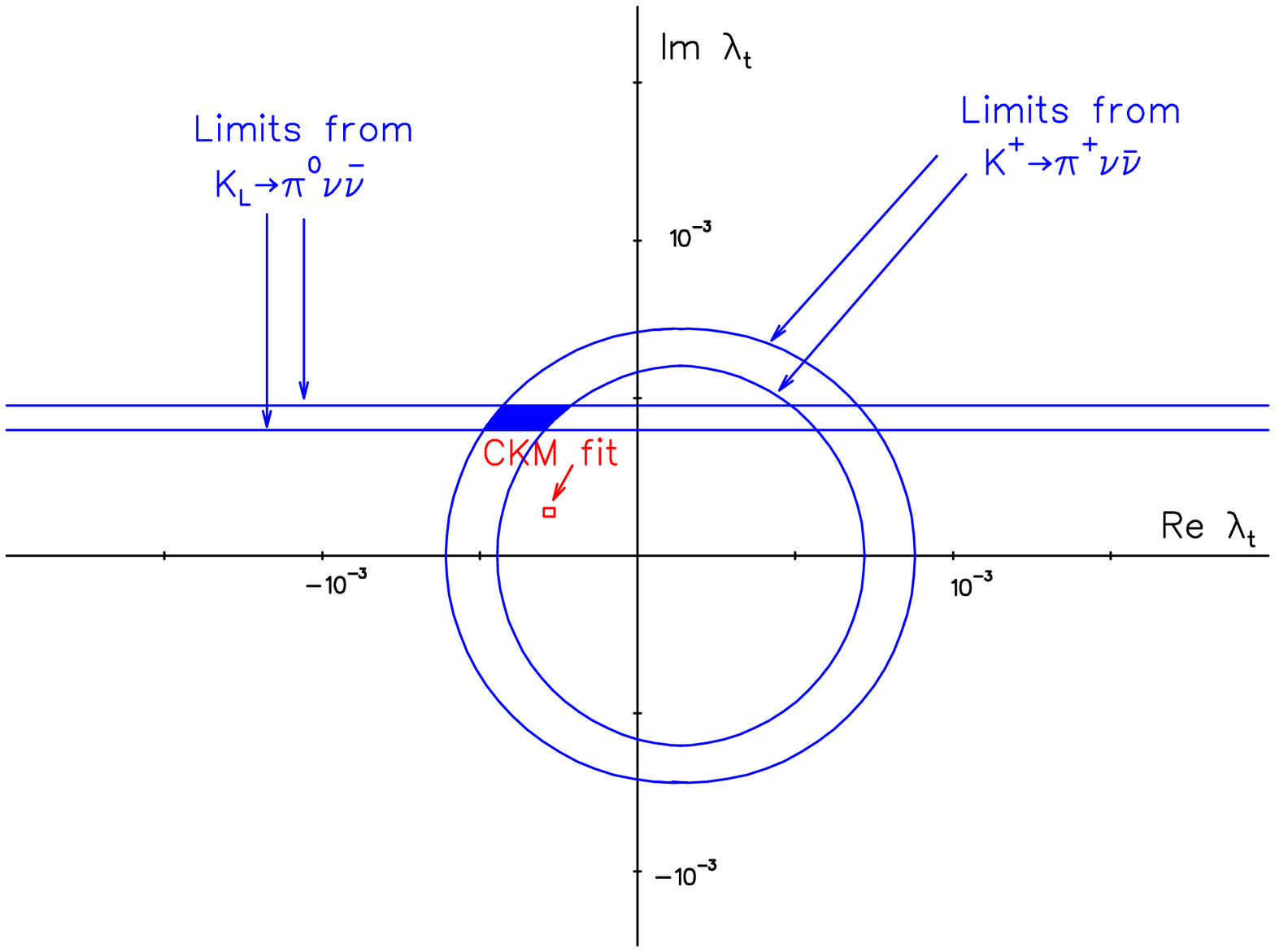}
  \caption{Left: Similar plot to Fig.~\ref{fig:ut90} after $10^{-11}$/event
$K^+ \to \pi^+ \nu\bar\nu$ experiment.  Assumes central value of
$B(K^+ \to \pi^+ \nu\bar\nu)$ stays the same and also that CKM fit contours
and precision on $m_t$ are improved by a factor 2.  Right: Similar
plot for possible scenario after 10\% measurements of $|\lambda_t|$
and $Im(\lambda_t)$.  Further improvements in CKM parameters and $m_t$
assumed.
    \label{fig:utfuture} }
\end{figure}

	Fig. \ref{fig:ut90} shows the 90\% CL constraints currently 
available
from \kmm~ and \kpnnp.  To extract a limit from the former we adopt the
value for the maximum long distance dispersive contribution from 
Ref. \cite{D'Ambrosio:1998jp}.  Also shown is the region in the $\lambda_t$
plane bounded by the CKM fit mentioned above \cite{Hocker:2001jb}. 
The two kinds of information are clearly consistent at the moment.
Note, however, that if the current central value of $B(K^+ \to \pi^+ 
\nu\bar\nu)$ should hold through E949, and expected progress is made in 
the $B$ 
sector, this agreement could prove short-lived, as shown on the left
of Fig.~\ref{fig:utfuture}.  When, eventually, we have 10-15\%  
measurements of \bkpnnp~and \bkpnn0, comparison of $K$ and $B$ 
results will become a critical test of the SM.   Fig.~\ref{fig:utfuture}
(right) illustrates a scenario in which such a failure is evident.

\section{Conclusions}

The success of lepton flavor violation experiments in reaching
sensitivities corresponding to mass scales of well over 100 TeV has
helped kill most models predicting accessible LFV in kaon decay.  The most
popular varieties of SUSY predict LFV at levels far beyond the current
experimental state of the art \cite{Belyaev:2000xt}.  Thus
new dedicated experiments in this area are unlikely in the near future.

        The existing precision measurement of \kmm\ will be very useful if
theorists can make enough progress on calculating the dispersive 
long-distance amplitude, perhaps helped by experimental progress in
$K_L \to \gamma \ell^+ \ell^-$, $K_L \to 4~leptons$, etc.  The exploitation
of \kmm~ would also be aided by higher precision measurements of
some of the normalizing reactions, such as $K_L \to \gamma\gamma$.

        \kpnnp\ will clearly be further exploited.  Two
coordinated initiatives are devoted to this: a $10^{-11}$/event experiment
(E949) just underway at the BNL AGS and a $10^{-12}$/event
experiment (CKM) recently approved for the FNAL Main Injector.  The first
dedicated experiment to seek \kpnn0~(E391a) is proceeding and an experiment (KOPIO)
at the AGS with the goal of making a $\sim 10\%$ measurement of
$Im(\lambda_t)$ is approved and in R\&D.

	Measurements of \kpnnp~and \kpnn0~can determine an
alternative unitarity triangle that will offer a critical comparison
with results from the $B$ system.  If new physics is in play in the
flavor sector, the two triangles will almost certainly disagree.

\begin{theacknowledgments}
I thank D. Bryman, M. Diwan, F. Gabbiani, D. Jaffe, S. Kettell, G. Valencia,
and M. Zeller for useful discussions, access to results, and other
materials. This work was supported by the U.S.  Department of Energy
under Contract No. DE-AC02-98CH10886.
\end{theacknowledgments}


\doingARLO[\bibliographystyle{aipproc}]
          {\ifthenelse{\equal{\AIPcitestyleselect}{num}}
             {\bibliographystyle{arlonum}}
             {\bibliographystyle{arlobib}}
          }
\bibliography{hf9}

\end{document}